\documentclass[prl,aps,floats,twocolumn]{revtex4}

\usepackage{amssymb} 
\usepackage{amsmath}
\usepackage{graphicx}
\usepackage{caption}
\usepackage{ctable}

\begin{document}

\title{Thermodynamic solution of the homogeneity, isotropy and flatness puzzles \\
 (and a clue to the cosmological constant)}

\author{Latham Boyle$^1$ and Neil Turok$^{1,2}$} 

\affiliation{$^{1}$Perimeter Institute for Theoretical Physics, Waterloo, Ontario, Canada, N2L 2Y5 \\
$^{2}$Higgs Centre for Theoretical Physics, University of Edinburgh, Edinburgh, Scotland, EH8 9YL}

\date{October 2022}
\begin{abstract}
We obtain the analytic solution of the Friedmann equation for fully realistic cosmologies including radiation, non-relativistic matter, a cosmological constant $\lambda$ and arbitrary spatial curvature $\kappa$.  The general solution for the scale factor $a(\tau)$, with $\tau$ the conformal time, is an elliptic function, meromorphic and doubly periodic in the complex $\tau$-plane, with one period along the real $\tau$-axis, and the other along the imaginary $\tau$-axis.  The periodicity in imaginary time allows us to compute the thermodynamic temperature and entropy of such spacetimes, just as Gibbons and Hawking did for black holes and the de Sitter universe.  The gravitational entropy favors universes like our own which are spatially flat, homogeneous, and isotropic, with a small positive cosmological constant.
\end{abstract}

\maketitle


\section{Introduction}

Soon after the discovery of black hole thermodynamics \cite{Bardeen:1973gs, Bekenstein:1973ur, Hawking:1974rv, Hawking:1975vcx}, Gibbons and Hawking \cite{Gibbons:1976ue} showed that one could elegantly compute the temperature and entropy of a general (charged, spinning) black hole, and of de Sitter (dS) spacetime, by noticing that these spacetimes are periodic in the imaginary time direction (see also the preceding work of Gibbons and Perry~\cite{Gibbons:1976es,Gibbons:1976pt}).  In this paper, we find the exact solution of the Friedmann equation for a general FRW universe including arbitrary amounts of radiation, non-relativistic matter (including baryons and dark matter), a cosmological constant and spatial curvature \cite{Kolb:1990vq, Mukhanov:2005sc}. Noticing that the scale factor $a(\tau)$ is periodic in the imaginary $\tau$ direction, we are able to compute the corresponding temperature and gravitational entropy.  Remarkably, the entropy obtained in this way favors universes like our own: homogeneous, isotropic and spatially flat, with small positive cosmological constant.

In earlier papers \cite{Boyle:2021jej, Turok:2022fgq}, we obtained similar results in the context of a simplified cosmology with radiation, a cosmological constant $\lambda$, and spatial curvature $\kappa$ but without non-relativistic matter. Our new result  strengthens the case that we have been developing \cite{Boyle:2018tzc, Boyle:2018rgh, Boyle:2021jej, Boyle:2021jaz, Turok:2022fgq, Boyle:2022lyw} for a simpler theory of the universe, not requiring inflation.

\section{General solution for $a(\tau)$}

The action for Einstein gravity coupled to matter is
\begin{equation}
  \label{Einstein_action}
  S=\int d^{4}x\sqrt{-g}\left(\frac{R}{2}-\lambda+{\cal L}_{matter}\right)
\end{equation}
where $\lambda$ is the dark energy (or cosmological constant). We shall use Planck units with $\hbar=c=k_B=8\pi G_{N}=1$ throughout. Now take the FRW line element
\begin{equation}
  ds^{2}=a^{2}(\tau)(-n^2 d\tau^{2}+\gamma_{ij}dx^{i}dx^{j})
\end{equation}
where $n$ is the lapse and $\gamma_{ij}$ is the metric on a maximally-symmetric 3-space of constant curvature $\kappa$; and take the matter to consist of radiation with energy density $r/a^4$ and non-relativistic matter (baryons and dark matter) with energy $\mu/a^{3}$, where $r$ and $\mu$ are positive constants.  Then the action becomes
\begin{equation}
  \label{mini_superspace_action}
  S=V_{3}\int d\tau\left[-3\frac{\dot{a}^{2}}{n}+n V(a)\right]
\end{equation}
where $V_{3}$ is the comoving spatial volume~\footnote{At fixed $\tau$, a closed ($\kappa>0$) universe is a 3-sphere with comoving volume $V_{3}=2\pi^{2}\kappa^{-3/2}$, while an open ($\kappa<0$) universe may have many different compact topologies -- see Ref.~\cite{ThurstonWeeks} -- with comoving  volume $V_{3}=\gamma \, 2\pi^{2}|\kappa|^{-3/2}$, with $\gamma \geq 1$. The pre-factor $\gamma$ generally grows as the topology of the negatively-curved 3-space becomes more complex. Since this $\gamma$-dependence is relatively unimportant for the considerations of this paper, in the plots we present here,  for simplicity we take $\gamma$ to be unity.}, $\dot{a}=da/d\tau$, and 
\begin{subequations}
  \begin{eqnarray}
    V(a)&=&-\lambda a^{4}+3\kappa a^{2}-\mu a-r \\
    &=&-\lambda(a^{4}-3\tilde{\kappa}a^{2}+\tilde{\mu}a+\tilde{r})
  \end{eqnarray}
\end{subequations}
where we have defined $\tilde{\kappa}\equiv\kappa/\lambda$, $\tilde{\mu}=\mu/\lambda$, $\tilde{r}=r/\lambda$. Varying with respect to $n$ (and then choosing the gauge $n=1$) yields the Friedmann equation
\begin{equation}
  \label{Friedmann_eq}
  3\dot{a}^{2}+V(a)=0
\end{equation}
while varying with respect to $a$ (and again taking $n=1$) yields the acceleration equation
\begin{equation}
  \label{2nd_Friedmann_eq}
  6\ddot{a}+V'(a)=0.
\end{equation}
First consider the critical ``Einstein static universe" (ESU) solutions, with constant scale factor $a_{esu}$: these require $\kappa>0$ and $\lambda>0$, with the parameters related by
\begin{equation}
  \label{ESU_cond}
    \frac{2\tilde{\kappa}^{3}}{\tilde{\mu}^{2}}=\frac{(1\!+\!8x)\!+\!(1\!+\!\tfrac{8}{3}x)\sqrt{1\!+\!\tfrac{32}{3}x}}{1+3\sqrt{1+\tfrac{32}{3}x}}\qquad(x\equiv\frac{\tilde{\kappa}\tilde{r}}{\tilde{\mu}^{2}}).
\end{equation}
Eq.~(\ref{ESU_cond}) defines an important boundary between different dynamical phases.  If the lhs of Eq.~(\ref{ESU_cond}) is greater than the rhs (which can only happen when $\kappa$ and $\lambda$ are both positive), we call it a ``turnaround" universe: its curvature $\kappa$ is sufficiently positive to cause a non-singular reversal from expansion to re-contraction, or contraction to re-expansion.  Otherwise ({\it i.e.},\ if $\lambda>0$ and $\kappa <\kappa_{esu}$, the positive, critical value set by Eq.~(\ref{ESU_cond})), the universe expands monotonically from the bang 
($a=0$) to the dS-like boundary ($a\to\infty$), or the reverse.

To find the general solutions for $a(\tau)$, first re-arrange the Friedmann equation (\ref{Friedmann_eq}) as 
\begin{equation}
da/\sqrt{-V(a)}=d\tau/\sqrt{3}.
\label{Friedmann_eq_diff}
 \end{equation}
Then, express $V(a)$ in terms of its four roots as
\begin{equation}
V(a)=-\lambda \prod_{i=1}^4 (a-a_{i}), 
\end{equation}
where $\{a_{1},a_{2},a_{3},a_{4}\}\equiv\{a_{++},a_{+-},a_{-+},a_{--}\}$ and
\begin{equation}
  \label{a_pm_pm}
  a_{\pm\pm}=\tfrac{1}{2}\big[(e_{\pm}^{})\pm\sqrt{-(e_{\pm}^{})^{2}-2\tilde{\mu}/(e_{\pm}^{})+6\tilde{\kappa}}\;\big],
\end{equation}
where we define $e_{\pm}\equiv \pm(2\tilde{\kappa} + z - P/z)^{1/2}$, with $z=(-Q+\sqrt{Q^{2}+P^{3}}\,)^{1/3}$, $P=-\tfrac{4}{3}\tilde{r}-\tilde{\kappa}^{2}$, and $Q=-\frac{\tilde{\mu}^{2}}{2}+\tilde{\kappa}^{3}-4\tilde{\kappa}\tilde{r}$; and on $a_{\pm\pm}$, the first $\pm$ subscript refers to $e_{\pm}$ while the second $\pm$ subscript is the sign in front of the square root in (\ref{a_pm_pm}). Now we define $a_{ij}\equiv a_{i}-a_{j}$, ${\rm sin}\,\varphi\equiv\sqrt{(a_{13}(a_{2}-a))/(a_{23}(a_{1}-a))}$,
\begin{subequations}
  \begin{eqnarray}
    m&=&\frac{a_{23}a_{14}}{a_{13}a_{24}}, \\
    \zeta&=&\frac{1}{2}\big[\frac{\lambda}{3}a_{13}a_{24}\big]^{1/2}.
  \end{eqnarray}
\end{subequations}
Then we integrate (\ref{Friedmann_eq_diff})  ({\it e.g.}, using Eq.~5 in Sec.~3.147 of Ref.~\cite{GradshteynAndRyzhik}) to obtain $F(\varphi,m)=\zeta\tau+C$, where $F(\varphi,m)$ is the elliptic integral of the first kind \cite{AbramowitzAndStegun}, and $C$ is a constant.  We invert this to obtain the Jacobi amplitude $\varphi={\rm am}(\zeta\tau+C,m)$.  Finally, taking ${\rm sin}$ of both sides and using ${\rm sin}({\rm am}(u,m))={\rm sn}(u,m)$, where ${\rm sn}$ is a Jacobi elliptic function \cite{AbramowitzAndStegun}, we obtain the general solution 
\begin{equation}
  a(\tau)=\frac{a_{2}a_{31}-a_{1}a_{32}{\rm sn}^{2}(\zeta\tau+C,m)}{a_{31}-a_{32}{\rm sn}^{2}(\zeta\tau+C,m)}.
\end{equation}
This solution is valid in general, even when the roots $a_{i}$ are complex.  Assuming $\lambda>0$, $C=0$ for a ``turnaround" universe and $C=\frac{i}{2}K(1-m)$ otherwise, with $K(m)$ the complete elliptic integral of the first kind \cite{AbramowitzAndStegun}. 

\begin{figure}
  \begin{center}
    \includegraphics[width=3.0in]{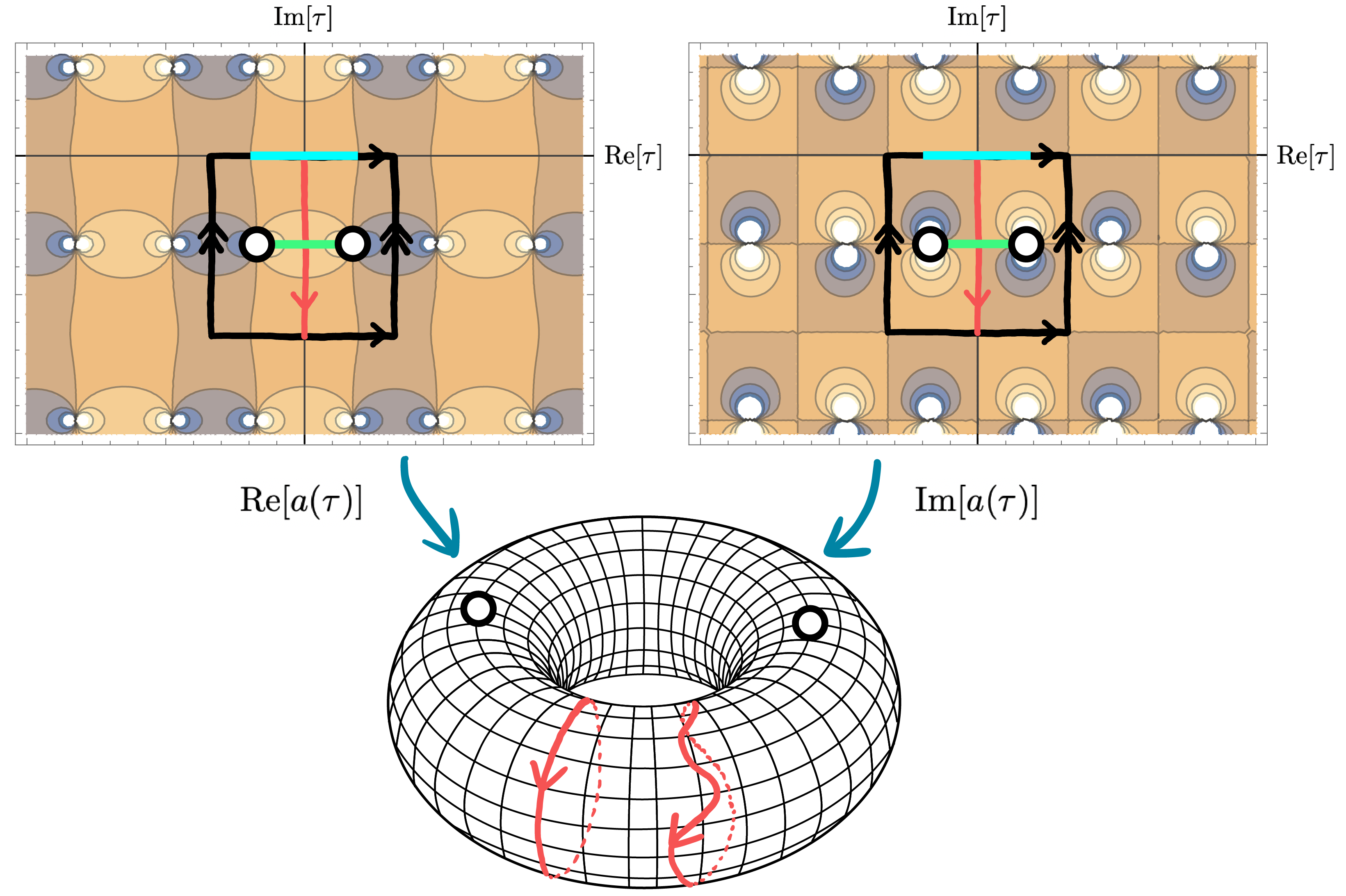}
  \end{center}
  \caption{Top: The real and imaginary parts of $a(\tau)$, with (black) rectangular fundamental domain. (For this example, $r=\tfrac{1}{4}$, $\mu=\tfrac{1}{2}$, $\kappa=1$, $\lambda=1$: a ``turnaround" cosmology.)  The thick horizontal blue (resp. green) line shows the Lorentzian solution along which the universe expands from a bang and re-collapses to a crunch, (resp. collapses from past dS infinity and re-expands to future dS infinity).   The vertical red line is the integration contour used to compute the gravitational entropy $\mathbf{S}_{g}$ and temperature $T_{g}$.  Bottom: We can regard $a(\tau)$ as a meromorphic function with two poles, on the torus obtained by identifying opposite edges of the fundamental domain. The red contour may be deformed to any topologically equivalent contour.}
\label{TorusFig}
\end{figure}

The general solution for $a(\tau)$ is thus an elliptic function -- meromorphic and doubly periodic in the complex $\tau$-plane, with one period oriented along the real $\tau$-axis and the other oriented along the imaginary $\tau$-axis.  As shown in Fig.~\ref{TorusFig}, these periods form the two sides of a rectangle, containing two poles of opposite residue; the solution $a(\tau)$ may be regarded as the rectangular tiling of the complex $\tau$-plane by copies of this rectangle. Alternatively, we can think of $a(\tau)$ as a meromorphic function on the torus formed by identifying the opposite edges of the rectangle (with the open circles indicating the two poles).  The imaginary $\tau$ period is 
\begin{equation}
  \label{Delta_tau}
  \Delta\tau=\pm 2i K(1-m)/\zeta.
\end{equation}

\section{Explicit Formulae for Gravitational Entropy ${\mathbf S}_{g}$ and Temperature $T_{g}$}

We can express the quantum amplitude to go from initial state $|i\rangle$ to final state $|f\rangle$ in time $\Delta t$ in two ways
\begin{equation}
  \label{amplitude}
  \langle f|{\rm e}^{-i H\Delta t}|i\rangle=\int{\cal D}[\varphi]{\rm exp}(iS[\varphi]),
\end{equation}
where the right-hand side is the usual path integral over all interpolating configurations $\varphi(x)$.  Now, by a standard argument \cite{Gibbons:1976ue}, if we identify the initial and final states $|f\rangle=|i\rangle$, sum over all states $|i\rangle$, and Wick rotate to Euclidean time $\Delta t = -i\Delta t_{E}=-i/T$, (\ref{amplitude}) becomes
\begin{equation}
  \label{partition_fn}
  \sum_{i}\langle i|{\rm e}^{-H/T}| i\rangle=\int{\cal D}[\varphi]{\rm exp}(-S_E[\varphi]),
\end{equation}
with $-S_E[\varphi]\equiv i S[\varphi]$, so the lhs -- the partition function at temperature $T$ -- may be expressed as the the amplitude to propagate along the Euclidean time direction $t_{E}$ by an amount $\Delta t_{E}=1/T$, and return to the initial state, while the rhs is the path integral over configurations periodic in Euclidean time, with period $\Delta t_{E}$.  

\begin{figure}
  \begin{center}
    \includegraphics[width=3.0in]{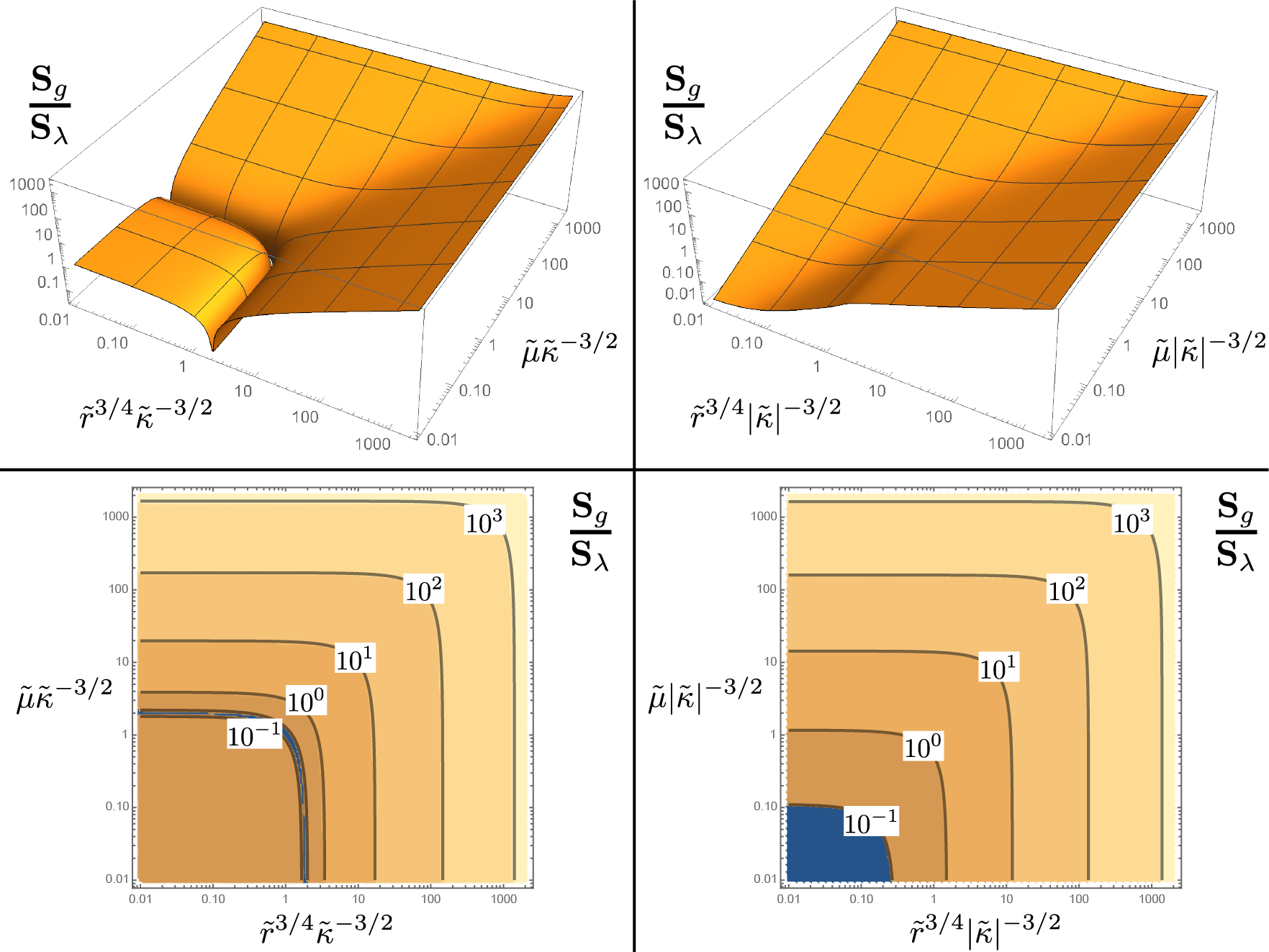}
  \end{center}
  \caption{The entropy ${\mathbf S}_{g}$ (\ref{entropy_formula}), relative to the de Sitter entropy ${\mathbf S}_{\lambda}=24 \pi^2/\lambda$, for $\kappa>0$ (left) and $\kappa<0$ (right), as a 3D plot (top) and contour plot (bottom).}
\label{EntropyFig}
\end{figure}

For a cosmological spacetime, the full (gravity plus matter) Hamiltonian $H$ vanishes due to time reparameterization invariance, so the lhs of (\ref{partition_fn}) just evaluates to the total number of states or, in other words, ${\rm exp}({\mathbf S}_{g})$, where ${\mathbf S}_{g}$ is the gravitational entropy.  On the other hand, just as in the black hole and de Sitter spacetimes \cite{Gibbons:1976ue}, the rhs may be evaluated in the semiclassical approximation, yielding $\sim {\rm exp}(-S_{E})={\rm exp}(iS)$, where $S$ is the action (\ref{mini_superspace_action}) for the classical spacetime, evaluated over a full period in imaginary time.  We conclude that 
\begin{equation}
  \label{entropy_integral}
  {\mathbf S}_{g}=i V_{3}\int_{0}^{\Delta\tau} d\tau\left[-3\dot{a}^{2}+V(a)\right],
\end{equation}
where the sign of $\Delta\tau$ in (\ref{Delta_tau}) must be chosen so that ${\mathbf S}_{g}>0$, since ${\rm exp}({\mathbf S}_{g})$ -- the number of microstates corresponding to a given macroscopic spacetime --  is $\geq1$.

The integration contour is depicted by the red curve in Fig.~\ref{TorusFig}: as shown there, it is a non-contractible loop winding once around the torus. Cauchy's theorem guarantees that the integral ${\mathbf S}_{g}$ is invariant under contour deformations which avoid the two poles.  Of all the topologically-distinct equivalence classes of contours we can choose, the physically correct contour is the one that sticks to the region where the scale factor $a$ (or more precisely its real part) remains large and positive (since the formula for the energy density of non-relativistic matter, $\mu/a^{3}$, may be only trusted as long as $a$ is neither too close to the bang nor negative).  When $\lambda>0$, there is always precisely one such equivalence class.

To evaluate (\ref{entropy_integral}), first imagine we are in the ``turnaround" case, so that $\kappa>0$, $\lambda>0$ and the roots $\{a_{1},a_{2},a_{3},a_{4}\}$
 are real, with $a_{1}>a_{2}>a_{3}>a_{4}$.  Using Eq.~(\ref{Friedmann_eq}), we can rewrite (\ref{entropy_integral}) as
\begin{equation}
  {\mathbf S}_{g}=4\sqrt{3} V_{3}\int_{a_{2}}^{a_{1}}da\sqrt{V(a)}.
\end{equation}
This integral may be evaluated by following the algorithm explained in Sections 13.1-13.8 of Ref.~\cite{Bateman1953Higher}.  The result is
\begin{equation}
  \label{entropy_formula}
   {\mathbf S}_{g}= \frac{\sqrt{3\lambda}\,V_{3}a_{23}}{2\sqrt{a_{13}a_{24}}}
  \!\left[C_{K}^{}K(\bar{m})+C_{E}E(\bar{m})+C_{\Pi}\Pi\Big(\frac{a_{12}}{a_{13}},\bar{m}\Big)\!\right]\!
\end{equation}
where $K(\bar{m})$, $E(\bar{m})$ and $\Pi(n,\bar{m})$ are the complete elliptic integrals of the 1st, 2nd and 3rd kinds, respectively, and 
\begin{subequations}
  \begin{eqnarray}
    C_{K}&=&a_{13}(a_{34}^{2}-a_{12}^{2}-\frac{4}{3}a_{14}a_{24}), \\
    C_{E}&=&8\tilde{\kappa}\frac{a_{13}a_{24}}{a_{23}}, \\
    C_{\Pi}&=&(a_{13}^{2}-a_{24}^{2})(a_{14}-a_{23}), \\
    \bar{m}&=&\frac{a_{12}a_{34}}{a_{13}a_{24}}.
  \end{eqnarray}
\end{subequations}
Although we derived this result for the ``turnaround" case where the roots $a_{1}>a_{2}>a_{3}>a_{4}$ are all real, the resulting formula (\ref{entropy_formula}) is valid in general, even when some or all of the roots are complex.  Our result for ${\mathbf S}_{g}$ is plotted in Fig.~\ref{EntropyFig}.  (In this paper, the horizontal axes are always labeled by $\tilde{r}^{3/4}|\tilde{\kappa}|^{-3/2}\sim {\mathbf S}_{r}/{\mathbf S}_{\lambda}^{3/4}$ and $\tilde{\mu}|\tilde{\kappa}|^{-3/2}\sim{\mathbf M}/{\mathbf S}_{\lambda}^{1/2}$, where ${\mathbf S}_{r}$ and ${\mathbf M}$ are the total radiation entropy and non-relativistic mass in the universe, respectively, and ${\mathbf S}_{\lambda}=24\pi^{2}/\lambda$ is the standard de Sitter entropy.)

As two checks on our formula, in the limit $\tilde{r}, \tilde{\mu} \downarrow 0$ with $\lambda$ and $\kappa$ positive, we recover the standard de Sitter entropy ${\mathbf S}_{\lambda}=24\pi^{2}/\lambda$. Second, all Einstein static universes, which are horizon-free, have ${\mathbf S}_{g}=0$. 

\begin{figure}
  \begin{center}
    \includegraphics[width=3.0in]{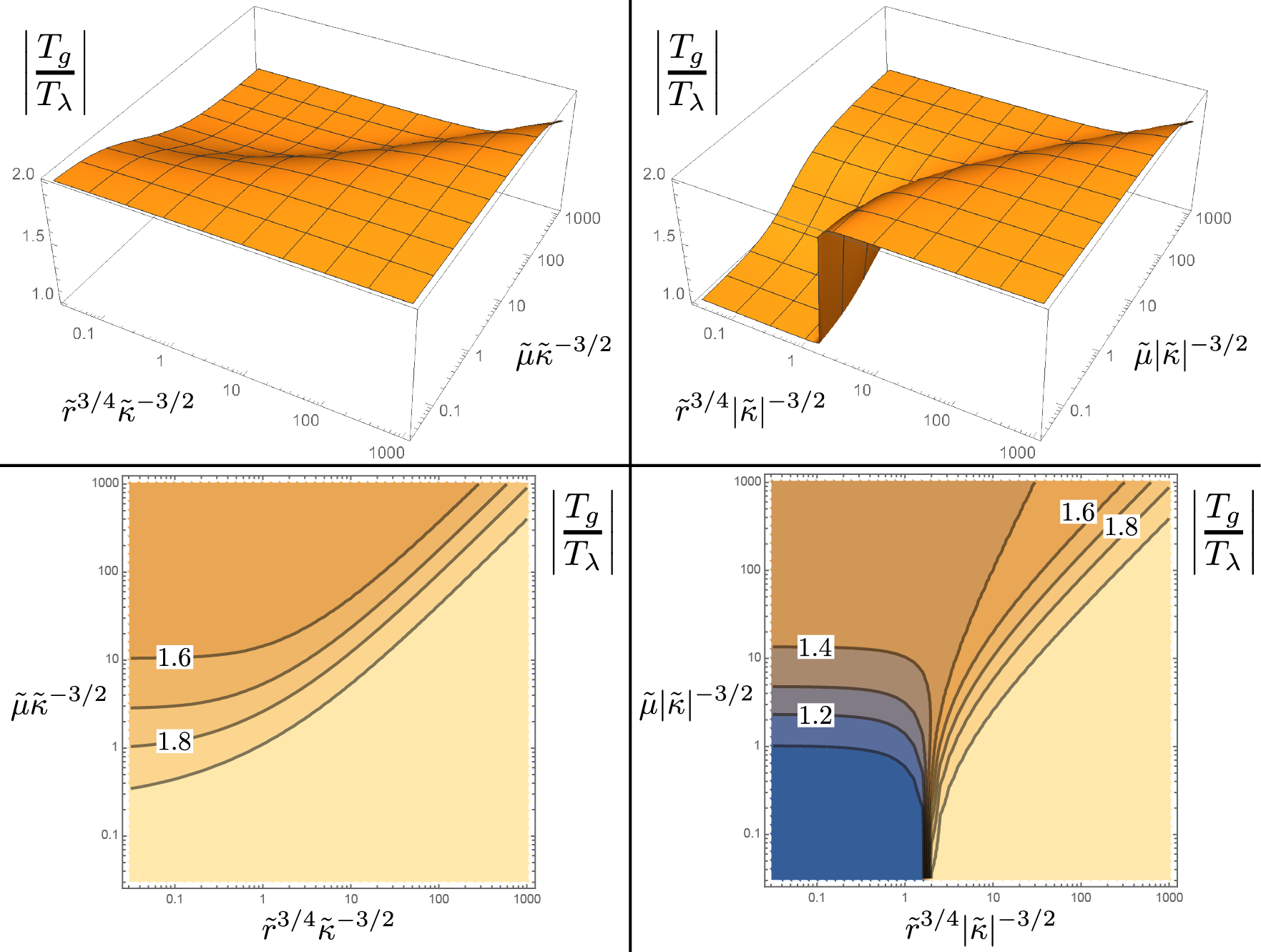}
  \end{center}
  \caption{Gravitational temperature $T_{g}$, relative to the dS temperature $T_{\lambda}= (2\pi)^{-1}\sqrt{\lambda/3}$, for $\kappa>0$ (left) and $\kappa<0$ (right), as 3D (top) and contour plot (bottom).}
\label{TempFig}
\end{figure}

The physical time $t=\int_{0}^{\tau} a(\tau)d\tau$ has three periods (one real and two imaginary) since, from Fig.~1, it depends on how many times the integration contour wraps around, or through, the hole in the torus, and also how it wraps around the poles in $a(\tau)$.  The first imaginary period, $\Delta t = \int_{0}^{\Delta\tau}a(\tau)d\tau$, evaluated along the same contour used to compute ${\mathbf S}_{g}$, yields a global gravitational ``temperature," illustrated in Fig.~\ref{TempFig}
\begin{equation}
  T_{g}\equiv i/\Delta t.
  \label{tglobal}
\end{equation}
Since the Euclidean geometry is not invariant under imaginary time translations, quantum field correlators defined in the Euclidean region will not be time translation invariant when they are continued to real time. As emphasized in Ref.~\cite{Turok:2022fgq}, we are describing an out of equilibrium ensemble. Hence, $T_g$ is a global quantity which is not locally measurable.  The second imaginary period, $\Delta t_{\lambda}=2\pi i R$ where $R=\pm \sqrt{3/\lambda}$ is the residue at the pole of $a(\tau)$,  yields the standard de Sitter temperature $T_{\lambda}=i/\Delta t_{\lambda}=(2\pi)^{-1}\sqrt{\lambda/3}$, which can be interpreted as the temperature of quantum fields as we approach the dS asymptopia (the pole of $a$)~\footnote{Note that, as $\tilde{\mu}$ and $\tilde{r}$ both tend to $0$, $T_{g}/T_{\lambda}$ tends to $2$ (where one might naively expect it to tend to $1$, since this is the dS limit).  The reason is topological: as long as either $\tilde{\mu}$ or $\tilde{r}$ are positive, the Euclidean solution has topology $S^{3}\times S^{1}$. It only converts to $S^{4}$ (the topology of Euclidean dS) when $\tilde{\mu}$ and $\tilde{r}$ are strictly zero.}.

Interestingly, the contour orientation needed to ensure ${\mathbf S}_{g}\geq 0$ implies that {\it $T_{g}$ is positive for a ``turnaround" universe, and negative otherwise.}  Negative temperatures can occur in systems with a finite number of accessible states \cite{Onsager, Ramsey}, as exemplified by the set of $\lambda>0$ universes we study with finite ${\mathbf S}_{g}$. The negativity of $T_{g}$ for universes like our own may be an important clue about how our macroscopic universe was born, as well as the microscopic ensemble that describes it.   

For $\lambda<0$, $T_{\lambda}=(2\pi)^{-1}\sqrt{\lambda/3}$ is imaginary, while $T_{g}$ diverges in the $\mu\to0$ limit and seems ill-defined for $\mu\neq0$ (since one cannot find a suitable contour where ${\rm Re}(a)$ remains positive), casting doubt on any thermodynamic interpretation of FRW cosmology with negative $\lambda$.

\section{Homogeneity, isotropy, flatness, $\Lambda$}

\begin{figure}
  \begin{center}
    \includegraphics[width=3.0in]{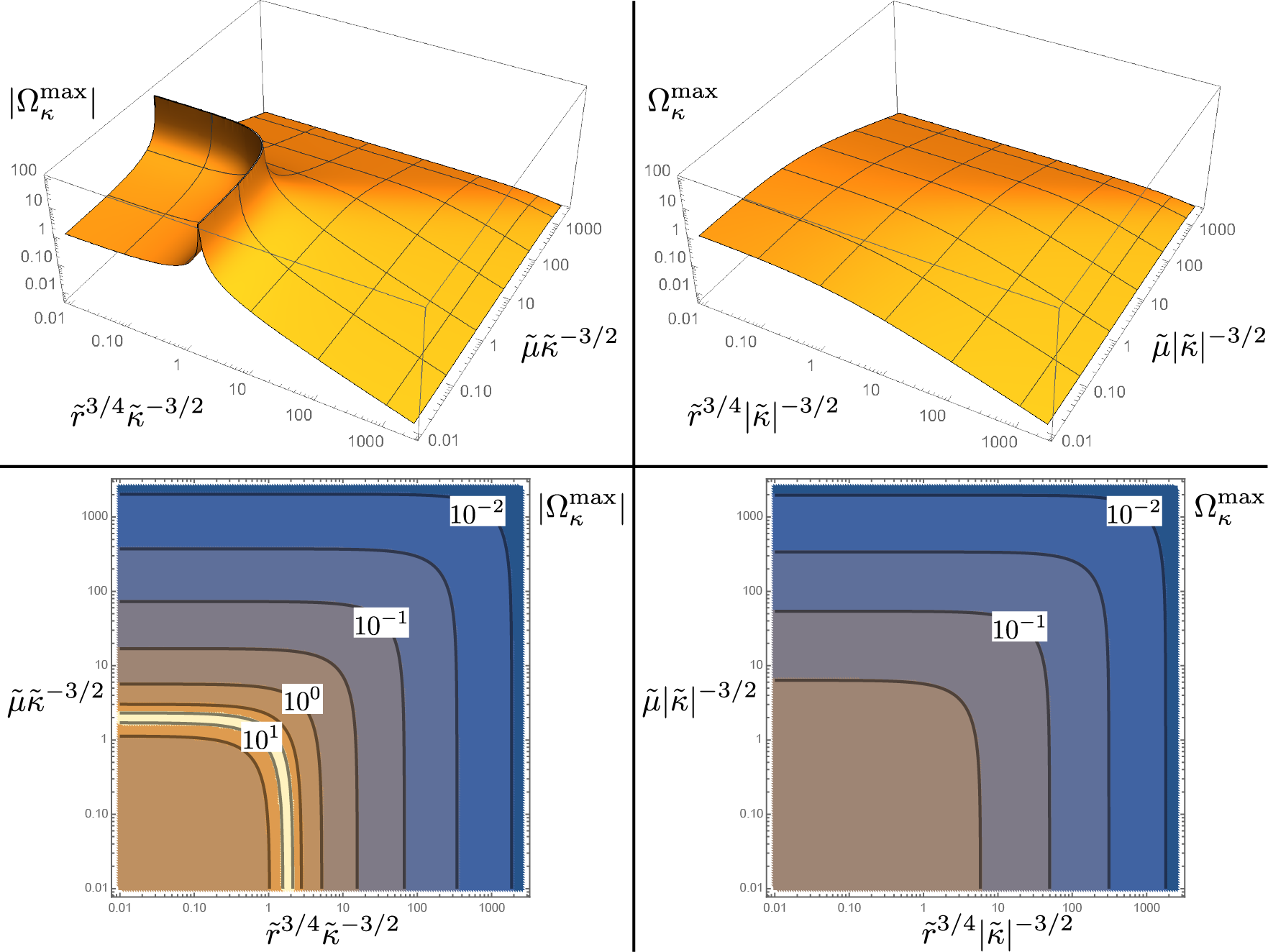}
  \end{center}
  \caption{The maximum curvature density $|\Omega_{\kappa}^{{\rm max}}|$ for $\kappa>0$ (left) and $\kappa<0$ (right), as a 3D plot (top) and contour plot (bottom).}
  \label{MaxCurvFig}  
\end{figure}

The ratio of curvature density to critical density is
\begin{equation}
  \Omega_{\kappa}(a)=-\frac{3\tilde{\kappa}}{a^{2}}\left[\frac{\tilde{r}}{a^{4}}+\frac{\tilde{\mu}}{a^{3}}-\frac{3\tilde{\kappa}}{a^{2}}+1\right]^{-1}.
\end{equation}
Setting the time derivative of this expression to zero, we see that $\Omega_{\kappa}$ reaches its maximal value when the scale factor satisfies 
$a=a_{\ast}$, where $a_{\ast}$ is the positive real solution of $a_{\ast}^{4}-\frac{1}{2}\tilde{\mu}a_{\ast}^{}-\tilde{r}=0$.  Thus, $\Omega_{\kappa}^{{\rm max}}=\Omega_{\kappa}(a_{\ast})$ where
$a_{\ast}=\tfrac{1}{2}(e_{\ast}+\sqrt{-e_{\ast}^{2}+\tilde{\mu}/e_{\ast}^{}})$, with $e_{\ast}=(z_{\ast}-P_{\ast}/z_{\ast})^{1/2}$, $z_{\ast}=(-Q_{\ast}+\sqrt{Q_{\ast}^{2}+P_{\ast}^{3}})^{1/3}$,
$P_{\ast}=(4/3)\tilde{r}$, and $Q_{\ast}=-(1/8)\tilde{\mu}^{2}$.  We plot $|\Omega_{\kappa}^{{\rm max}}|$ in Fig.~\ref{MaxCurvFig}.

Comparing Fig.~\ref{EntropyFig} and Fig.~\ref{MaxCurvFig}, note that the contours have the same shape, and as ${\mathbf S}_{g}$ increases $|\Omega_{\kappa}^{{\rm max}}|$ decreases.   Universes with entropy above some threshold ${\mathbf S}_{g}$ are universes with maximum curvature below some threshold $|\Omega_{\kappa}^{{\rm max}}|$; and increasing the threshold ${\mathbf S}_{g}$ decreases the threshold $|\Omega_{\kappa}^{{\rm max}}|$.  In other words, the most entropically likely universes are those (like our own) in which the curvature never becomes significant throughout cosmic history.  This solves the flatness problem.

To go beyond this ``zeroth order" result, next add small inhomogeneities ({\it i.e.}\ tensor and scalar perturbations, $h$ and $\zeta$) and anisotropies ({\it i.e.}\ tensor perturbations with wavelengths longer than the Hubble radius) to these preferred nearly-flat backgrounds.  To quadratic order, these are described by the actions $S_{\zeta}\!\sim\!\int\! d^{4}x\,z^{2}(\zeta'{}^{2}\!-\!c_{s}^{2}(\nabla\zeta)^{2})$ for scalar perturbations and $S_{h}\!\sim\!\int\! d^{4}x\,a^{2}(h'{}^{2}\!-\!(\nabla h)^{2})$ for tensor perturbations \cite{MFB}.  As seen above, to ensure the leading order answer ${\mathbf S}_{g}$ is positive, the integration contour in this (and any non-turnaround) case must run {\it up} the imaginary $\tau$ axis; and, as explained in Refs.~\cite{Boyle:2018tzc, Boyle:2021jej}, the perturbations $h$ and $\zeta$ satisfy reflecting boundary conditions at the bang that ensure they are even functions of $\tau$, and hence are real along the imaginary $\tau$ axis.  Together these facts imply $S_{h}^{(2)}$ and $S_{\zeta}^{(2)}$ both contribute {\it negatively} to ${\mathbf S}_{g}$ \cite{Turok:2022fgq}.   So for fixed values of the conserved quantities characterizing the cosmology ({\it e.g.} $r$ and $\mu$ or, more physically, the total entropy in radiation ${\mathbf S}_{r}\sim r^{3/4}V_{3}$ and the total mass in non-relativistic matter ${\mathbf M}\sim \mu V_{3}$), inhomogeneities and anisotropies {\it decrease} ${\mathbf S}_{g}$, so are entropically disfavored.

So far, we have treated $\lambda$ as a fixed constant, but now consider it as another parameter in the ensemble \cite{Henneaux:1989zc}.
For $\lambda<0$, the thermodynamic interpretation is suspect, and there is no reason to expect a large gravitational entropy.  However, for $\lambda>0$, the entropy ${\mathbf S}_{g}$ is huge, and increases as we decrease $\lambda$.  For example, for reasonably flat universes ($|\Omega_{\kappa}^{{\rm max}}|\ll1$) we have ${\mathbf S}_{g}\sim {\mathbf S}_{r}\lambda^{-1/4}$ when $(4\tilde{r}/9)^{1/2}\gtrsim(\tilde{\mu}/2)^{2/3}$ and ${\mathbf S}_{g}\sim {\mathbf M}\lambda^{-1/2}$ when $(4\tilde{r}/9)^{1/2}\lesssim(\tilde{\mu}/2)^{2/3}$.  From this we see that, for fixed ${\mathbf S}_{r}$ and ${\mathbf M}$, ${\mathbf S}_{g}$ increases as $\lambda$ approaches zero from above, and the highest entropy ${\mathbf S}_{g}$ is achieved in the limit $\lambda\to 0^{+}$.  In other words, the entropy ${\mathbf S}_{g}$ also seems to favor universes (like our own) with a tiny positive $\lambda$. Our result echoes and extends to a realistic universe those of \cite{Baum:1983iwr, Hawking:1984hk, Coleman:1988tj}. We will discuss the associated statistical ensemble in a forthcoming paper. 

{\bf Acknowledgements:}  NT is supported by the STFC Consolidated Grant `Particle Physics at the Higgs Centre' and the Higgs Chair of Theoretical Physics. Research at Perimeter Institute is supported by the Government of Canada, through Innovation, Science and Economic Development, Canada and the Province of Ontario through the Ministry of Research, Innovation and Science.


\begin{thebibliography}{99}
\bibitem{note1}
We only consider cosmologies in which space is compact.
\end{thebibliography}
\end{document}